\documentclass[12pt,a4paper]{elsart}

\usepackage{graphicx}
\usepackage{fancyhdr}
\usepackage{amssymb}
\usepackage{pslatex}

\begin{document} 

\begin{frontmatter}

\title {Energy of Ni/Ni$_{3}$Al Interface: a Temperature-Dependent Theoretical Study}

\author{A. S. Martins\corauthref{asm}},
\corauth[asm]{martins@if.uff.br}
\author{M. F. de Campos}

\address{Departamento de F\'\i sica, ICEx, Universidade Federal Fluminense}

\begin{abstract}
Using plane wave {\textit{ab-initio}} and Monte Carlo techniques, the Ni/Ni$_{3}$Al interfacial energy is studied and the results from the different techniques are critically compared. Two issues deserved special attention: the dependency of the interface energy with the supercell size, for the {\textit{ab-initio}} calculations, and the temperature dependence of the interface energy. The calculations show that from 0K up to 1000 K that energy decreases by 1 mJ/m$^2$. 
\end{abstract}

\begin{keyword}
DFT \sep Monte Carlo \sep Superalloys \sep Interfacial Energy

\end{keyword}
\end{frontmatter}

\section{Introduction}

Superalloys is a general term used to describe a myriad of high-performance materials which possess high mechanical resistance at high temperatures. In particular, the Ni-Al alloys have several technological applications as rocket motors and turbine blades. Coherent Ni3Al precipitates are one of the main strengthening mechanisms in these family of alloys. An important characteristic influencing the precipitation and growing of the Ni$_{3}$Al particles is the Ni/Ni$_3$Al or $\gamma/\gamma'$ interface energy. In a recent review by Ardell \cite{ardell2011}, the author present a re-evaluation of the available data on coarsening  of $\gamma'$ precipitates at the light of the (TIDC) and (LSW) theories of coarsening, aiming to provide the best possible values of interfacial free energies $\sigma$ of $\gamma/\gamma'$ interfaces. Estimates of the $\gamma/\gamma'$ interface energy by theoretical and experimental methods indicates that its value is between 10-70 mJ/m$^2$\cite{ardell2011}. A previous theoretical estimation using embedded atom method (EAM) at 0 K temperature \cite{farkas1993} resulted in 22 mJ/m$^2$ value. An important question often neglected is the influence of the temperature on the interface energy. This question will be addressed in the present study. 

Theoretical methods based on quantum-mechanics are always desirable for the computation of material properties and, with the advent of the density functional theory (DFT), full quantum-mechanical (\textit{ab-initio}) calculations up to thousands of atoms become possible by using massive parallel computation. Although the number of atoms seems modest, important structural and thermodynamical material properties can be accurately obtained. However, the usual calculations using DFT rarely treats systems with sizes greater than few hundreds of atoms. On the other hand, classical simulations like Monte Carlo allow to handle systems with some thousands of atoms, depending upon the potential employed as model of the interaction between the atoms: many-body potentials requires a bigger computational effort compared to the simple pair-potentials.

In this article, DFT and $NPT$ Monte Carlo simulations of the Ni/Ni$_{3}$Al interface are reported and their results critically compared. The thermal behavior of the system is carried out within the harmonic approximation for the atomic vibrations and the temperature dependence of the interfacial energy is obtained adding to the DFT/Monte Carlo energies the contributions from the phonons. The article is organized as follows: the next two sections are devoted, respectively, for presenting the theoretical formalism and preliminary calculations and the remaining sections focus on the determination of the interfacial energy for Ni/Ni$_{3}$Al. The conclusions are left to the last section.

\section{Methodology}  

The DFT calculations were in the spin-polarized formalism as implemented in the Quantum Espresso\footnote{Quantum ESPRESSO is an integrated suite of computer codes for electronic-structure calculations and materials modeling at the nanoscale. It is based on density-functional theory, plane waves, and both norm-conserving and ultrasoft pseudopotentials.} package \cite{giannozzi2009}, employing Vanderbilt Ultrasoft pseudopotentials \cite{vanderbilt1990} for the core electrons of the Ni and Al elements. The calculations were carried out within the generalized gradient approximation (GGA), with PBE parametrization for exchange correlation potential among the electrons \cite{pbe}.

In order to complement the DFT study, Monte Carlo simulations in the isothermal-isobaric $NPT$ ensemble were too performed, with the Cleri-Rosato potential \cite{cleri1993} employed as a model for the atom-atom interaction. The Monte Carlo codes employ the Metropolis importance sampling algorithm to perform the stochastic evolution of the system. In Monte Carlo simulations the system's energy corresponds only to the configurational part of the real system, that is, the kinetic part is not included, although the Metropolis sampling guarantee the constance of the temperature. The physical properties of the system in MC are calculated as ensemble averages of their instantaneous values for each configuration generated in phase space, one configuration being defined as the instantaneous set of atomic positions.

The functional form of the Cleri-Rosato potential, based on the second-moment approximation of a tight-binding hamiltonian of the transition metals, is built in order to yield a good description of the cohesive energy, elastic constants and thermal behavior of the transition metals. The total energy of the system is given by
\begin{equation}
E = \sum\limits_{i} \sum\limits_{i \neq j} \left\{ A_{\alpha\beta}e^{-p_{\alpha\beta}\left(\frac{r_{ij}}{d_{\alpha\beta}} - 1\right)} -  \sqrt{\xi^{2}_{\alpha\beta} e^{-2p_{\alpha\beta}\left(\frac{r_{ij}}{d_{\alpha\beta}} - 1\right)}} \right\},
\label{crpot}
\end{equation}
where $\alpha$ and $\beta$ indicate the atomic specie and $r_{ij}$ the distance between the $i$ and $j$ atoms, with the indices $i$ and $j$ running over the all atoms of the system. For all calculations presented in this article, the potential parameters $A_{\alpha\beta}$, $p_{\alpha\beta}$, $\xi_{\alpha\beta}$ and $q_{\alpha\beta}$ were taken from the reference \cite{cleri1993}.

In order to take into account the temperature effect on the DFT/MC energies, a thermostatistical analysis \cite{amaral01,amaral02} was performed on the relaxed structures. The calculation is carried out in two steps: the phonon frequencies are determined from the relaxed configuration within the harmonic approximation, by diagonalizing the corresponding dynamical matrix (in the $\Gamma$ point) and, from the vibrational spectrum, the thermodynamical data were obtained from the vibrational partition function. Denoting by $U_0$ the total energy of the DFT/MC calculations, the internal energy is given by
\begin{equation}
E_{int} = U_{0} + \frac{1}{2}\sum\limits_{i=1}^{3N}\hbar\omega_i + \sum\limits_{i=1}^{3N}\frac{\hbar\omega_i}{e^{\beta\hbar\omega_i}-1},
\label{eneint}
\end{equation}
where $\beta = 1/KT$ and the summation index $i$ runs over all vibrational modes. The two first terms in \ref{eneint} correspond to the zero point energy and the last one is the vibrational energy.

\section{Preliminary Studies}

The magnetism of the TM atoms comes from the unfilled $3d$ shell. On the other hand, bulk aluminum is not magnetic. In DFT calculations, it is necessary to assess previously the quality of the pseudopotentials and establish the size of the plane wave basis and the number of $k$ points used in the integration over the Brillouin zone, in order to guarantee a good convergence of the total energy. As the interest here is quantifying the interfacial energy of the Ni/Ni$_{3}$Al, convergence studies on the above issues were performed, by checking two particularly situations: the resulting bulk magnetism of the Ni, Al and Ni$_{3}$Al and their equilibrium lattice parameters $a_0$: their calculated values are shown in the table \ref{table:tab1}. It can be noticed that the used pseudopotentials were able to reproduce correctly the magnetization and the equilibrium lattice parameters of the bulk materials. it is worth noting the excellent agreement when comparing them with the experimental values reproduced in the table.

\begin{table}[ht]
\caption{\label{table:tab1}: Magnetic moment per atom $M$ and optimal lattice parameters of the bulk $Ni$ and $Ni_{3}Al$, with the corresponding experimental values.}
\centering
\begin{tabular}{|c|c|c|c|c|}
\hline \hline
\textbf{Compound} & $M$ ($\mu_B$) & $M_{exp}$ ($\mu_B$) & $a_0$ ($\textsc{\AA}$) & $a_{0}^{exp}$ ($\textsc{\AA}$)  \\
\hline
Ni  & 0.67  & 0.7 & 3.53  & 3.52 \\
Al  & 0.00 &  0.0 & 4.05  & 4.05   \\
Ni$_{3}$Al & 0.00 & 0.0   & 3.57 & 3.57 \\
\hline \hline
\end{tabular}
\end{table}

In all calculations presented in the Table \ref{table:tab1}, the cutoff energy was set to 50 Ry and a $k$-point Monkhost-Pack grid of $8 \times 8 \times 8$ was employed. For the interface energy calculations, the cutoff energy was set to 40 Ry. It is known that ultrasoft pseudopotentials allow using smaller cutoff energy compared to norm conserving pseudopotentials \cite{martins1991} and the presented calculations indicate that this value is enough to describe the compounds: for the bulk Ni, for example, the difference between the total energies is 0.02 $eV$ when the cutoff energy is decreased from 50 Ry to 30 Ry.

With respect to the Monte Carlo Simulations, the quality of the material description is related to the used potential. As mentioned before, the parameters of the Cleri-Rosato potential were determined in order to result in a good description of the cohesive energy, elastic constants and thermal behavior \cite{cleri1993}. However, the potential does not take into account the magnetic interaction between the atoms.

\section{The Ni/Ni$_{3}$Al Interface}

\begin{figure}[ht]
\begin{center}
\includegraphics[height=8cm]{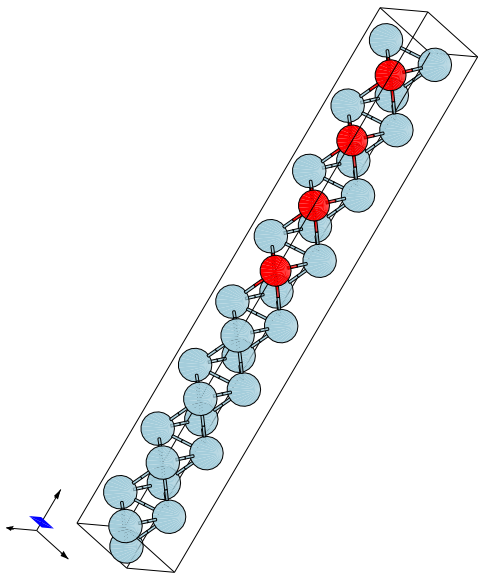}
\caption{A 32-atom supercell: the red balls correspond to the Al atoms. \label{fig1}}
\end{center}
\end{figure}

For both DFT and MC calculations, orthorhombic supercells with dimension of $a=n_z\cdot a_0$, $b=n_y\cdot a_0$ and $c=n_x\cdot a_0$ were generated, where $n_z$, $n_y$ and $n_x$ correspond, respectively, to the number of FCC unit cells along the $z$, $y$ and $x$ directions\footnote{As the $\gamma^{'}$-Ni$_{3}$Al and $\gamma$-Ni lattice parameters are close, the value of lattice parameter $a_0$ for all systems can be taken, for example, as a arithmetic average of the heir values.}. For all generated Ni/Ni$_{3}$Al supercells, it was considered only the (001) interface and the number of Ni unit cells was equal to the number of Ni$_3$Al ones ($n_z$ is always even). As an example of one of considered structures, Fig. \ref{fig1} shows a 32-atom supercell ($n_z = 8$)

Due to the computational cost of the DFT calculations, the dimensions of the supercells along $x$ and $y$ were assigned to one lattice parameter, $n_x=n_y=1$, that is, $b = c$. Along the $z$ direction, the number of FCC unit cells was varied in order to get a set of energy values, which will be a function of the of the length of Ni$_{3}$Al slabs, $Z_{alloy}$: the thermodynamical limit of the interface energy corresponds to $Z_{alloy}\rightarrow\infty$. All DFT calculations were spin polarized and employed Monkhost-Pack grid of $8 \times 8 \times 4$. 

In the Monte Carlo calculations, the supercells dimensions were $n_x = n_y = 3, 4$ and $n_z = 8$. In this way, it was possible to check the convergence of the interface energy with the interface area. In all MC simulations, an equilibration phase of 15.000 MC steps were performed, and the physical quantities computed correspond to averages over the configurations generated along 10.000 MC steps in the production phase. For the thermodynamic analysis, the internal and free energies, the entropy and the specific heat correspond to averages over a total of 50 configurations. The temperature in the MC calculation was set to 1 K, just to allow a small relaxation of the atomic positions, and the pressure was assigned to zero.

\begin{figure}[ht]
\begin{center}
\includegraphics[height=8cm]{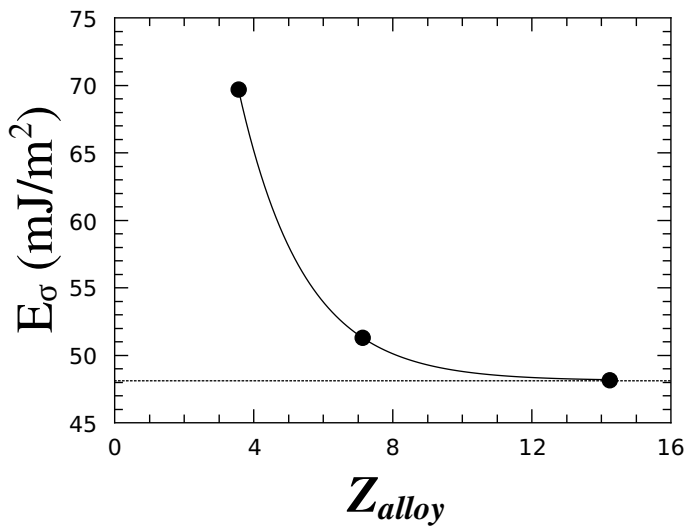}
\caption{Interface energy of Ni/Ni$_{3}$Al as function of the length of the Ni$_{3}$Al alloy along the $z$ direction. The full and dotted lines correspond, respectively, to the \textit{ansatz} mentioned in the text and the extrapolated value $E_{\infty}$ from it. \label{fig2}}
\end{center}
\end{figure}

The interface energy is calculated in two steps: (\textit{i}) The total energy of the Ni/Ni$_3$Al supercell is calculated with full atomic and volum relaxation, $E_{tot}(a, b, c)$, with $a$, $b$, and $c$ representing the relaxed lattice parameters. (\textit{ii}) The total energies of the single $\gamma$ or $\gamma^{'}$ are calculated for supercells with initial dimensions equal to the previous step, but fixing $b$ and $c$ and allowing $a$ to be relaxed. The interface energy is given by:

\begin{equation}
E_{\sigma} = \frac{\left\lbrace E_{tot}(a, b, c) - \frac{1}{2} \left[E_{\gamma}+E_{\gamma^{'}} \right] \right\rbrace}{2S},
\label{itfcene}
\end{equation}
where $S$ is the surface area, $S=b\times c$. It is expected that the interface energy depends upon the supercell size along the $z$ direction: an small value of $a$ implies a significant interaction between the Ni/Ni$_{3}$Al interface and their periodic images. On the other hand, for a greater $a$ value the Ni atoms far from the surface remain themselves to their bulk positions, as desired. In Fig. \ref{fig2} we have the plot of the interface energy $E_{\sigma}$ as function of the lenght of the Ni$_{3}Al$ along $z$, $Z_{alloy}$. It can be noticed a very fast convergence of $E_{\sigma}$: assuming that $E_{\sigma}$ follows the \textit{ansatz} $E_{\sigma} = E_{\infty} + A e^{-Z/a}$ the extrapolated value is $E_{\infty} = 48.1$ $mJ/m^2$.

According with Fig. \ref{fig2}, the extrapolated value of $E_{\sigma}$ is within the experimental range of $10-70$ $mJ/m^2$ expected for this interface \cite{ardell2011}. In the theoretical context, classical atomistic simulations using many body interatomic potential developed by Voter \cite{voter1987} found a value of 22 $mJ/m^2$ \cite{farkas1993} and DFT calculations by Silva and colaborators \cite{silva2007} assign the value of $39.6$ $mJ/m^2$ for the $(001)$ interface.

Following the same methodology, the interface energies for the MC calculations were $E_{\sigma}=42$ mJ/m$^2$ for the 288 atom supercell ($n_x=n_y=3$, $n_z = 8$) and 40 mJ/m$^2$ for the 512 atoms supercell ($n_x=n_y=4$, $n_z = 8$). Both values are in a good agreement with the DFT calculations. The difference can be essentially attributed to the non-inclusion of the magnetic interaction in the MC calculations, and less importantly, to the size of the interface area.

\section{Thermodynamic Analysis of the Ni/Ni$_3$Al interface}

\begin{figure}[ht]
\begin{center}
\includegraphics[height=8cm]{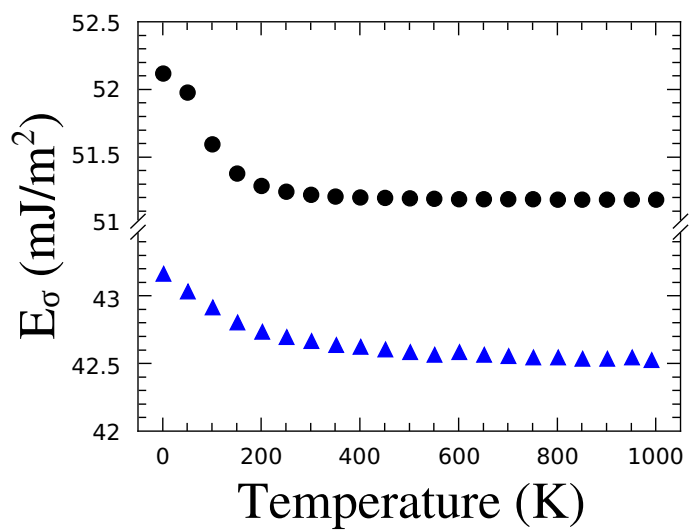}
\caption{Interface energy of Ni/Ni$_{3}$Al as function of the temperature for both DFT (circles) and MC (triangles) techniques. \label{termo}}
\end{center}
\end{figure}

From the previous section, the methodology proposed here yield an interface energy within the range of the acceptable experimental values. However, as claimed by Ardell \cite{ardell2011}, the DFT calculations are performed at 0 K, and theoretical evaluations have not been reported concerning the temperature effect on the interface energy. Following the methodology outlined in the Section 2, in the figure \ref{termo}, the surface energy presented as function of the temperature, for both DFT and MC. We can notice that the temperature effect is modest: the surface energy decreases around 1 mJ/m$^2$ as the temperature goes to very high values. According to a recent report by Ardell \cite{ardell2011}, a difference of the interface energy of 2.81 mJ/m$^2$ and 3.38 mJ/m$^2$ when the temperature is raised from 898 K to 988 K (see Tables 5 and 6). Beside the discrepancy between the employed models, both result in a more significantly reduction of the interface energy with the temperature.

The surface energy was calculated adding to the DFT/MC energy the contribution of the phonons and the vibrational energies, according Eq. \ref{eneint}. Although the individual values of the Ni, Ni$_3$Al and Ni/Ni$_3$Al energies show a significant decrease in their values, the interface energy, as defined in Eq. \ref{itfcene}, depends upon the difference among their values and, thus, the decrease of the interface energy was smaller compared to the individual contributions. In addition, it can be noticed in Fig. \ref{termo}, both DFT and MC predictions converge asymptotically to well defined values as the temperature goes to higher values.

The modest reduction of the interface energy reported here cannot be attributed to the anharmonic effects, because the calculations were performed far from the melting point, around of 1500 K for the Ni and Ni$_{3}$Al. On the other hand, the model employed for computing the interface energy, equation \ref{eneint}, does not take into account the effect of the variation of the volume with the temperature. Thus, maybe the agreement could be improved adding the effect of the volume dilatation on the internal energy.

\section{Conclusions}

In this article, DFT and Monte Carlo simulations of the Ni/Ni$_{3}$Al interface are reported and their results critically compared. The interface energies around 0 K determined in both methodologies agrees very well with the differences attributed to the non inclusion of the magnetic interaction in MC and, less significantly, to the small interface area in the DFT calculations. The thermal behavior of the system is obtained within the harmonic approximation for the atomic vibrations and the temperature dependence of the interface energy is obtained adding the contributions from the phonons to the DFT/Monte Carlo energies. Apart the significant decrease of the internal energies with respect to the temperature for the Ni, Ni$_3$Al and Ni/Ni$_3$Al systems, the interface energy shows a corresponding small variation, $\approx$ 1 mJ/m$^2$, for both DFT and MC calculations. The modest variation is attributed to a kind of the cancellation of the temperature contribution to the interface energy of the Ni/Ni$_{3}$Al and the former Ni and Ni$_{3}$Al systems. The addition of the contribution of the energy due the volume variation due the temperature can result in a more realistic reduction of the interface energy with the temperature.

\end{document}